\documentclass[ reprint, onecolumn, superscriptaddress, floatfix, amsmath,amssymb, aps ]{revtex4-2}
\usepackage[utf8]{inputenc}
\usepackage[letterpaper,top=2cm,bottom=2cm,left=3cm,right=3cm,marginparwidth=1.75cm]{geometry}

\usepackage[colorlinks=true, allcolors=blue]{hyperref}
\usepackage{amssymb}
\usepackage{amsmath}
\usepackage{amsthm}

\usepackage{subcaption}
\usepackage{graphicx}
\usepackage{xcolor}
\usepackage{bm}

\def \Xv{\boldsymbol{X}}
\def \Yv{\boldsymbol{Y}}
\def \Tv{\boldsymbol{T}}

\def \yv{\boldsymbol{y}}
\def \Fv{\boldsymbol{F}}

\newcommand{\bea}{\begin{equation} \begin{aligned}}
\newcommand{\eea}{\end{aligned} \end{equation} }

\newcommand{\eqn}[1]{Eq.~\eqref{#1}}

\begin{abstract}
Metasurfaces and complex photonic components are increasingly co-designed with computational back-ends via end-to-end optimization, yet such optimizations are expensive and opaque---obscuring the role of the optics and any fundamental performance limits. We develop two information-theoretic objectives, based on Shannon capacity and Fisher information, that isolate the photonic contribution to image formation. Both are closed-form, data-free functions of the transfer matrix, requiring no training data, and yield designs whose reconstruction quality matches end-to-end optimization. We prove that for both objectives, and for a broader family with a shared mathematical structure, the optimal transfer matrix is a permutation matrix: each source’s emission is concentrated on a single, distinct detector, a condition we call generalized focusing.
This holds regardless of source/detector geometry, as we demonstrate in settings where conventional imaging intuition offers no guidance, including a two-way imager, imaging through a random scattering medium, and Hermite--Gauss mode sorting. The root of this constraint is an ``intensity bottleneck:'' nonnegative intensity measurements admit only Kronecker deltas as a complete orthonormal basis. We further show that this bottleneck, and the generalized-focusing optimum, persist for coherent and partially coherent sources---the constraint is the detector array, not the source coherence.
\end{abstract}

\begin{document}

\title{End-to-end meta-imagers: \\Information-theoretic objectives and generalized focusing optima}

\author{Lukas Kienesberger}
\affiliation{Department of Physics, Yale University, New Haven, CT, USA}
\affiliation{Energy Sciences Institute, Yale University, New Haven, CT, USA}
\author{Zeyu Kuang}
\affiliation{Institute for Theoretical Physics, TU Wien, Vienna A-1040, Austria}
\author{Yaxi Liu}
\affiliation{Energy Sciences Institute, Yale University, New Haven, CT, USA}
\affiliation{Department of Applied Physics, Yale University, New Haven, CT, USA}
\author{Owen D. Miller}
\email[Corresponding author: ]{owen.miller@yale.edu}
\affiliation{Department of Physics, Yale University, New Haven, CT, USA}
\affiliation{Energy Sciences Institute, Yale University, New Haven, CT, USA}
\affiliation{Department of Applied Physics, Yale University, New Haven, CT, USA}

\maketitle

\section{Introduction}
Computation is playing an increasingly central role in imaging~\cite{mait2018computational,delbracio2021mobile}. At the same time, metasurfaces and other complex photonic components~\cite{yu2014flat,khorasaninejad2016metalenses,chen2020flat,rotter2017light} offer the possibility of dramatically new, and potentially more informative, optical point-spread functions that are fed to advanced computational back-ends. The powerful tandem of complex photonics and back-end computation is ideally jointly designed, with ``end-to-end'' optimization recently showing state-of-the-art performance and capabilities~\cite{stork2008theoretical,chakrabarti2016learning,sitzmann2018end,chang2018hybrid,tseng2021neural,lin2021end,lin2022end,arya2024endtoend,wetzstein2020inference,barbastathis2019deep}. Yet end-to-end optimizations can be expensive, if averaging is done over many random source instances and noise samples. Perhaps more importantly, they are opaque~\cite{Lalanne2026-vy}: it can be difficult to discern the role that the optics plays in the image-formation process, and what (if any) fundamental limits might constrain their ultimate performance.

In this paper, we use information theory to propose two deterministic, data-free metrics for the optical/photonic part of a computational ``meta-imager'' (generally conceived). We focus on the prototypical case in which all detectors measure intensities only. Both metrics are closed-form functions of the transfer matrix alone, requiring no training data or noise samples; optimizing them and then separately training the computational back-end produces performance comparable to full end-to-end optimization, as we verify in a variety of examples. The key benefit is that the objectives enable theoretical analysis of optimal point-spread functions. We find a simple, general criterion: the optimal incoherent transfer matrix is a \emph{permutation matrix}---each source's light emission must be spatially focused to a single, distinct detector, a condition we call ``generalized focusing.'' We prove this for both proposed objectives and a larger class satisfying positivity and convexity constraints, regardless of the spatial arrangement of sources and receivers, and demonstrate it computationally in geometries where conventional imaging intuition offers no guidance, including imaging through a random scattering medium. The root of this constraint is the ``intensity bottleneck'': intensity measurements collapse signals onto a space in which the only orthogonal vectors are Kronecker deltas. This bottleneck is independent of source coherence, and we show that the generalized-focusing optimum persists for coherent and partially coherent inputs alike.

\begin{figure}[tb]
  \centering
  \includegraphics[width=0.8\linewidth]{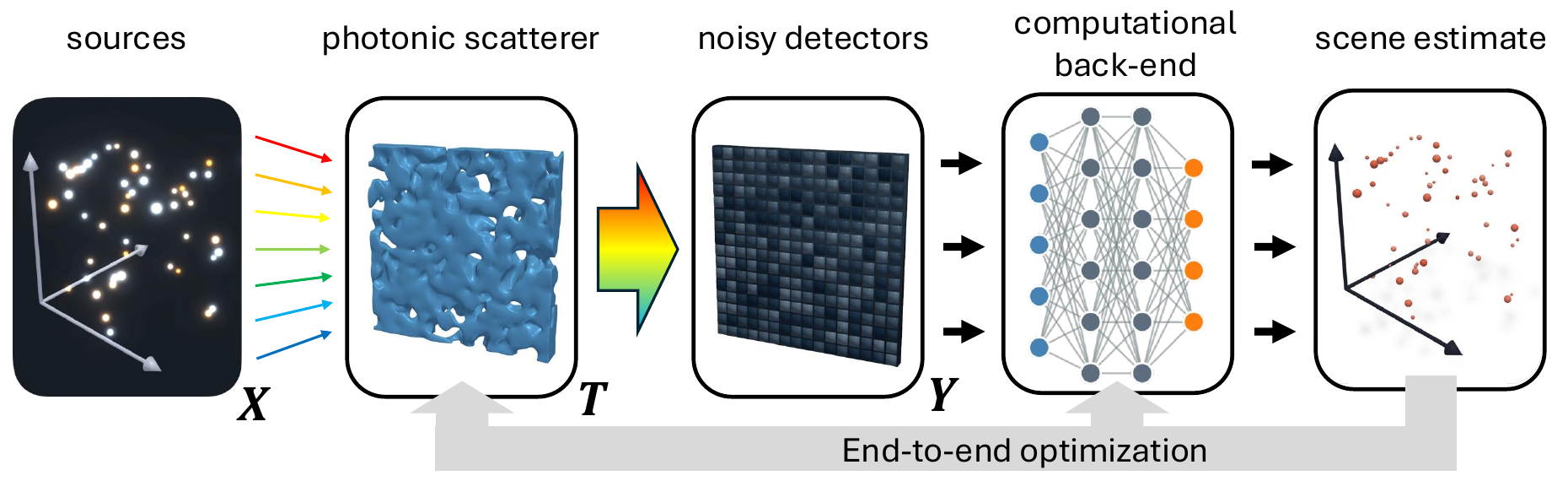}
  \caption{End-to-end computational imaging pipeline: sources ($\Xv$) emit through a photonic scatterer (transfer matrix $\Tv$) onto noisy intensity detectors ($\Yv$), followed by a computational back-end that produces a scene estimate. End-to-end optimization (gray bracket) jointly tunes the scatterer and back-end but is expensive and opaque. We develop information-theoretic objectives for the photonic stage alone---from $\Xv$ through $\Tv$ to $\Yv$---that are data-free, closed-form, and reveal that the optimal $\Tv$ is a permutation matrix (generalized focusing).}
  \label{fig:fig1}
\end{figure}
Classical computational imaging has largely taken near-diffraction-limited, lens-like optics as given and focused on algorithms for reconstructing images from noisy, band-limited measurements~\cite{mait2018computational}. Deconvolution uses a known point-spread function to invert residual aberrations or blur, effectively ``correcting'' imperfect but otherwise lens-like optics. Coded-aperture imaging~\cite{fenimore1978coded,levin2007image} engineers variably obstructed pupil patterns so that the collective response---across the detector, and in multi-shot variants across exposures---encodes more of the scene than a single unobstructed aperture would. Wavefront coding trades sharpness for a PSF deliberately invariant to a nuisance parameter such as defocus, with the induced blur removed computationally~\cite{dowski1995extended}. Super-resolution methods complement these approaches by extracting subdiffraction detail from multi-frame data or strong scene priors~\cite{park2003super,yang2010image,dong2016image}, exploiting information that a single-shot PSF does not directly transmit. 
More recently, the scope has expanded beyond conventional refractive and diffractive optics to metasurfaces and other alternative photonic systems whose point-spread functions can differ dramatically from lens-like responses~\cite{colburn2018metasurface,tseng2021neural,bayati2022inverse,chakravarthula2023thin}, and whose response is increasingly co-designed with the reconstruction algorithm~\cite{lin2021end,lin2022end,arya2024endtoend}. A common figure of merit used to compare such systems and guide their design is the condition number (or, more generally, the singular-value spread) of the discretized transfer matrix, motivated by the intuition that well-conditioned systems yield more stable reconstructions~\cite{barbastathis2019deep}. Yet the condition number is both discretization-dependent---its value depends on the chosen source/detector basis and resolution---and difficult to connect directly to specific optical or photonic response.

Information theory offers a natural route to general computational imaging objectives. Its origins in imaging systems extend almost as far back as Shannon, with work by Fellgett and Linfoot~\cite{fellgett1955assessment,linfoot1955information} enabling physically motivated, discretization-independent ``pixel counts'' that have proven useful to this day. (This connects nicely with the separate electromagnetic pursuit of quantifying the number of independent degrees of freedom an imaging system can transmit under diffraction, sampling, and noise constraints~\cite{toraldo1969degrees,huck1999information,ashok2003information,miller2019waves}). A recent thread, pursued especially by the Waller group, revives information theory as a direct optimization objective, fitting probabilistic noise models to measurements and maximizing the resulting mutual information jointly with the optics~\cite{pinkard2024information,markley2025ideal,kabuli2026lensless}, with an emphasis on realistic statistical modeling. Another active thread addresses the two-point (Rayleigh) resolution problem, where Fisher information has been used to characterize how precisely the separation between two incoherent point sources can be estimated, and to exhibit measurement designs that beat Rayleigh's criterion~\cite{tsang2016quantum,nair2016farfield,lupo2016ultimate,paur2016achieving,tham2017beating,tsang2019resolving,wadood2025super}. Those results are sharp but largely specific to the two-point geometry and do not obviously extend to scene reconstruction with many sources in arbitrary arrangements. Our analysis targets that latter case: we take information theory as a design objective for many-source intensity-based imaging (possibly with inverse-designed scatterers and nonstandard source/detector arrays), and use it to derive a universal condition---generalized focusing---on the optimal photonic response.

\section{Information-theoretic photonic objectives}
In this section, we develop two objectives intended to isolate the contributions of the photonic components, including detectors, to the image formation process (for ``optimal'' back-end processing, suitably defined). We start with notational definitions and then develop two objectives, one from a Shannon capacity perspective and the second from a Fisher-information-based, image reconstruction perspective.

We assume $N$ incoherent sources with intensities arranged in a source vector $\Xv$, producing intensities $\Yv$ measured at $N$ detectors. (In Sec.~\ref{sec:Discussion} we discuss extensions to unequal numbers of source and detector points.) We make no assumptions on the geometrical arrangement of sources and receivers (and consider nonstandard arrangements below). The optical system (including free space, metasurfaces, lenses, etc.) is characterized by its ($N \times N$) incoherent point spread function $\Tv$:
\begin{align}
    \Yv = \Tv \Xv.
    \label{eq:iPSF}
\end{align}
Typically, images are normalized pixelwise between 0 and 1, which corresponds to lower and upper bounds on each element of $\Xv$: $0\le X_n\le P$ for all $n$, with $P=1$. Many of our results do not rely heavily on the normalization of the sources/emitters, which could also, for example, be subject to a total power ($\ell_1$ norm) constraint: $\sum_{n=1}^N X_n\le P_{\mathrm{tot}}$.

The linearity of Eq.~\ref{eq:iPSF} relies on the emitters being perfectly incoherent; full or partial coherence at the emission side would imply a nonlinear input--output relation. We assume incoherence in the theoretical information analysis and examples below to leverage the linear (point spread function) formalism. Yet, as we discuss in Sec.~\ref{sec:bottleneck}, it is really the intensity measurement at the output side that constrains optimal response, not source coherence: we show there that non-mixing transfer is optimal for any level of source coherence, provided the detectors measure only intensities.

Noise makes $\Yv$ a random vector. We assume independent additive Gaussian readout noise~\cite{janesick2007photon,healey1994radiometric}, $\boldsymbol{\varepsilon}$ with standard deviation $\sigma$ and mean zero (for sufficiently small $\sigma$ values that one need not worry about overall intensities going negative), giving
\begin{equation}
    \Yv = \Tv \Xv + \boldsymbol{\varepsilon},
    \qquad
    \boldsymbol{\varepsilon}\sim\mathcal N(\boldsymbol{0},\sigma^2 \boldsymbol{I}_N).
    \label{eq:InputOutput}
\end{equation}
The measured output $\Yv$ becomes a multivariate Gaussian with conditional probability distribution:
\begin{equation}
    p(\yv| \Xv)
    = (2\pi\sigma^2)^{-N/2}\exp\!\left[-\frac{1}{2\sigma^2}\|\yv-\Tv\Xv\|_2^2\right].
    \label{eq:gaussian_likelihood}
\end{equation}
The central question is then: how does $\Tv$ affect the image formation process, and can one identify optimal $\Tv$ matrices? We take two information-theoretic perspectives.

\paragraph*{Shannon capacity.}
In the first, we view the photonic system as an information channel~\cite{cover2006elements,telatar1999capacity}. A natural starting point is the mutual information $I(\Xv;\Yv)$, the reduction in uncertainty about the source $\Xv$ that comes from observing the measurement $\Yv$. For a given noise model, maximizing $I(\Xv;\Yv)$ jointly with the optics is a powerful design technique~\cite{pinkard2024information,markley2025ideal,kabuli2026lensless}. Here we instead ask: over all possible source distributions $p(\Xv)$, what is the maximum information rate the photonic channel can support? This is the \emph{channel capacity}, $C = \sup_{p(\Xv)} I(\Xv;\Yv)$, a distribution-free figure of merit for the optical front-end. Related channel capacities have been extensively explored in MIMO communications~\cite{foschini1998limits,telatar1999capacity}, albeit typically for coherent communications in fixed or random (but not designable) environments. Our physical setting (intensity-only, nonnegative inputs) has previously been considered in free-space optical communications with LEDs and photodetectors~\cite{lapidoth2009capacity,li2020capacity}. Under a high-SNR approximation with additive Gaussian noise (standard deviation $\sigma$) and maximum per-emitter power $P$, the capacity reduces to (cf.\ SM for a self-contained derivation)
\begin{equation}
C \approx \log\!\left(\frac{|\det(P\Tv)|}{(\sqrt{2\pi e}\,\sigma)^N}\right).
\label{eq:CapacityUnreg}
\end{equation}
To handle ill-conditioned systems where some singular values of $\Tv$ fall below the noise floor, we use the regularized surrogate
\begin{equation}\label{eq:ShannonCapacity}
C = \frac{1}{2}\log\det\!\left(\boldsymbol{I}+\frac{P^2}{2\pi e\sigma^2}\,\Tv^T\Tv\right),
\end{equation}
which recovers Eq.~\ref{eq:CapacityUnreg} for well-conditioned $\Tv$ and contributes zero for sub-SNR modes (cf.\ SM). In the high-SNR limit, the input distribution achieving the capacity is approximately uniform, so \eqn{eq:CapacityUnreg} is nearly achieved by generic, evenly distributed scenes. This equation is derived in Ref.~\cite{Moser2017-wi} for free-space communication rates; here, we propose that it has alternative utility for optimizing and understanding imaging systems.

\paragraph*{Fisher information / CRB.}
Complementary to the capacity viewpoint, one can regard imaging as parameter estimation: given noisy measurements, how accurately can one reconstruct source intensities? From this perspective, one might infer the unknown ``parameter'' $\Xv$ from (random-variable) measurements $\Yv$ by maximizing the likelihood of measuring $\Yv$ over all possible $\Xv$ values. Central to the maximum-likelihood framework is the Fisher information matrix~\cite{barrett2004foundations}, 
\begin{align}
\Fv(\Xv)=\left\langle\nabla_{\Xv}\log p(\Yv|\Xv)\,\nabla_{\Xv}\log p(\Yv|\Xv)^T\right\rangle,
\end{align}
which is the covariance of the ``score'' (itself the derivative of the log-likelihood function). A small Fisher information (in any singular value) implies a small curvature of the log-likelihood function, in which case noisy measurements may impede high-fidelity parameter recovery (since many parameter values may have similar likelihoods). Conversely, large Fisher information implies better parameter-recovery feasibility. These intuitions are made mathematically precise by the \emph{Cram\'er--Rao bound} (CRB), which says that the covariance (uncertainty) of \emph{any} unbiased estimator of the parameter $\Xv$ is bounded below by the inverse Fisher matrix,
\begin{equation}
\mathrm{cov}_{\Xv}(\widehat{\Xv})\succeq\Fv(\Xv)^{-1},
\label{eq:crb_X}
\end{equation}
in the matrix positive semidefinite sense. Large Fisher information enables better parameter estimation; small Fisher information prevents it. Fisher information has been applied extensively in optical imaging to bound single-emitter localization precision through a given (typically diffraction-limited) PSF~\cite{thompson2002precise,ram2006beyond,chao2016fisher}, and more recently as an objective for PSF engineering in single-emitter localization~\cite{pavani2009three,shechtman2014optimal}; here we apply it to full-scene reconstruction, with the entire transfer matrix designable. For the Gaussian noise model of Eq.~\ref{eq:gaussian_likelihood}, the Fisher information takes a simple form, 
$\Fv(\Xv) = \Tv^T\Tv / \sigma^2$. A well-defined objective must be single-valued; we choose to take the trace of the estimator's covariance matrix, thereby representing the total expected squared error. We denote the trace of the inverse of the Fisher information matrix as $\mathcal{E}$:
\begin{equation}
\mathcal E = \sigma^2\,\operatorname{Tr}[(\Tv^T\Tv)^{-1}].
\label{eq:FisherUnreg}
\end{equation}
The same metric has recently been used as a data-free objective for inverse design of computational spectrometers~\cite{yu2025spectrometer}, and a related metric---the sum of inverse singular values $\sum 1/\sigma_j$ rather than inverse squared---has been proposed in a similar context~\cite{ma2026inverse}. As with the capacity objective, \eqn{eq:FisherUnreg} diverges for ill-conditioned $\Tv$; regularizing yields the surrogate
\begin{equation}
\mathcal{E} = \sigma^2\,\operatorname{Tr}\left[\left(\Tv^T\Tv+\frac{1}{\mathrm{SNR}^2}\boldsymbol{I}_N\right)^{-1}\right].
\label{eq:FisherObjective}
\end{equation}
where $\mathrm{SNR} \approx P/\sigma$ sets the crossover between measurement-dominated and noise-dominated modes, so that sub-SNR modes contribute a bounded penalty rather than diverging. (Physically, this just corresponds to the fact that one should not try to estimate parameters whose signal at the detectors is below the noise level.) In electromagnetic scattering, the falloff of singular values is typically sharp---an exponential or quasi-exponential decay associated with the crossover between propagating and evanescent waves~\cite{miller2019waves,kuang2025bounds,miller2025tunneling}---so the regularized objectives effectively select the well-conditioned (or nearly so) channels and optimize over those.

Both objectives are closed-form functions of $\Tv$ alone, requiring no training data or noise samples, and as we verify in Sec.~\ref{sec:CompExamples}, they produce photonic designs competitive with end-to-end optimization. Before turning to those examples, we show that both objectives (and a broader class with similar structure) share a simple optimum: the best $\Tv$ is always a permutation matrix.

\section{Generalized focusing}
\label{sec:GenFoc}
\begin{figure}[tb]
  \centering
  \includegraphics[width=0.6\linewidth]{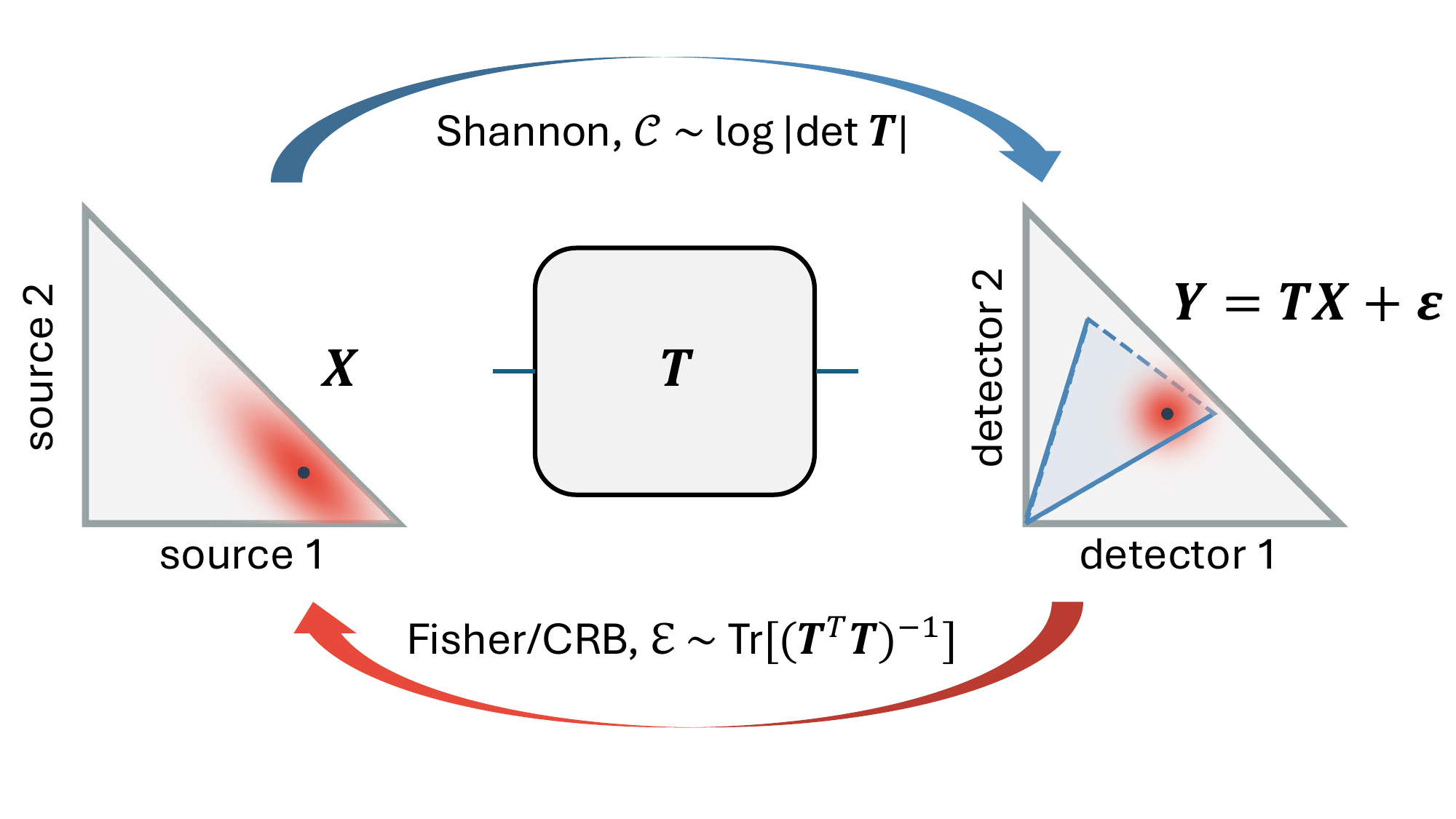}
  \caption{Geometric interpretation of the two information-theoretic objectives, illustrated for $N=2$ sources and detectors. The source excitation $\Xv$ (black dot) lies in the input simplex (gray triangle, left). The transfer matrix $\Tv$ maps this to a measurement $\Yv = \Tv\Xv + \boldsymbol\varepsilon$ (black dot, right) within the signal simplex (blue triangle), whose volume is set by $|\!\det \Tv|$; additive noise (red cloud) perturbs $\Yv$. The Shannon capacity (blue arrow, forward direction) maximizes the ratio of signal volume to noise volume. The Fisher/CRB metric (red arrow, backward direction) minimizes the reconstruction uncertainty: noise in the measurement maps back to an error region in the source space (red cloud, left) whose size is bounded by $\operatorname{Tr}[(\Tv^T\Tv)^{-1}]$.}
  \label{fig:FSInterp}
\end{figure}

The mathematical expressions above can be understood more physically with the geometric illustration in Fig.~\ref{fig:FSInterp}. Consider first sources with a total-power constraint ($\sum X_i \leq P$, an $\ell_1$-norm constraint): any incoherent excitation across the $N$ sources will lie in the ``unit simplex;'' in two dimensions (i.e., for two source points), it will lie in the gray triangle at the left of Fig.~\ref{fig:FSInterp}. The $N\times 1$ source excitation $\Xv$ passes through the photonics, leading to detector intensities prescribed by $\Tv \Xv$, with $N \times N$ incoherent point spread function matrix $\Tv$, which therefore must have columns whose entries are nonnegative and sum to less than 1. Hence the columns of $\Tv$ all lie in the positive orthant, excitations with unit power are convex combinations of these columns (the dashed line segment on the right-hand side of Fig.~\ref{fig:FSInterp}), and excitations with powers between 0 and 1 fill the interior of this region, comprising the shaded blue simplex (triangle, in 2D). The black dot within this region represents $\Yv = \Tv \Xv$ for the single $\Xv$ (black dot on the left). The additive Gaussian noise perturbs $\Yv$ in a circularly symmetric manner around this point, as represented by the red density. (If instead each emitter's amplitude is individually bounded, $0 \leq X_i \leq P$, the input set is a hypercube rather than a simplex and its image is a parallelepiped, but the conclusions below are unchanged: the parallelepiped volume is $N!$ times the simplex volume, so it is maximized by the same $\Tv$.)

The Shannon perspective of Eq.~\ref{eq:CapacityUnreg} can be viewed as a statement of maximizing ``forward information flow'' through the process. The term $|\det(P\Tv)|$ is the volume of the parallelepiped formed by the columns of $\Tv$ (scaled by excitation power $P$), which is a simple constant multiple of the blue simplex volume that represents the volume of the space of possible (noise-free) measurements. The denominator in Eq.~\ref{eq:CapacityUnreg} is proportional to the volume of a noise ball. Their ratio gives then a typical Shannon count of independent signals that can be transmitted without error, with the base-2 logarithm then converting this to a ``bitrate.''

The Fisher perspective, of Eqs.~(\ref{eq:FisherUnreg},\ref{eq:FisherObjective}), corresponds conversely to optimizing ``backward flow,'' starting with a measurement $\Yv$ and reconstructing $\Xv$ as closely as possible. Just as a single, noise-less measurement is ideally transformed back to the correct initial excitation point, the noise ball in the measurement space will be transformed to a noise volume in the input space. By the Cramer--Rao bound of Eq.~\ref{eq:crb_X}, the eigenvalues of the covariance matrix are bounded below by the inverses of the eigenvalues of $\Tv^T \Tv$: hence the noise ball on the right must map to a domain containing the ellipsoid with semi-axis given by the inverse eigenvalues of $\Tv^T \Tv$ (red blob on the left). Small singular values of $\Tv$, of course, lead to large uncertainty in the reconstruction.

These geometric pictures then immediately offer intuition about the optimal $\Tv$ matrices to maximize our information-theoretic objectives: \textbf{the optimal $\Tv$ matrix is a permutation matrix}. In the Shannon picture, one can see that to maximize the volume of the blue simplex on the right-hand side, the columns of $\Tv$ (which form the edges of the simplex emanating from the origin) should align with the coordinate axes. The identity matrix is one option, but any permutation is equivalent: the volume will be identical for any ordering of these vectors. Similarly, in the Fisher picture, the volume of the noise ball is minimized when all eigenvalues of $\Tv$ have magnitude 1 and $\Tv$ is orthonormal. ($\Tv$ has real-valued entries, so one can specialize from unitary to orthonormal.) Since an orthonormal matrix with all nonnegative entries must be a permutation matrix (as no two columns can have a nonzero entry at the same row index), the same conclusion follows. 

Physically, an optimal permutation-matrix $\Tv$ means that every source should scatter its light to a single, distinct detector---a condition we call \emph{generalized focusing}. The ``generalized'' qualifier is important: the permutation need not respect any spatial ordering, so \emph{any} source can map to \emph{any} detector regardless of their geometric arrangement. The focusing condition stands in contrast to coherent systems, where mode sorters~\cite{morizur2010programmable,labroille2014efficient,fontaine2019laguerre,miller2013self} can implement any unitary transformation and the optimal transfer can be realized in any complete orthonormal basis. With intensity-only detection, the nonnegativity of the measurements restricts which orthonormal bases are accessible: the only nonnegative vectors that can form a complete orthonormal basis are the Kronecker deltas. This is the root of the generalized-focusing constraint---not a limitation of the optics or the reconstruction algorithm, but a fundamental property of the nonnegative measurement space.


One can widen the class of objectives, beyond the Shannon and Fisher objectives, for which permutation matrices are optimal. In particular, every function of the singular values of the form $F(\lambda_1,\dots,\lambda_N) = \sum_j f(\lambda_j^2)$ with $f$ smooth, strictly convex, and monotonically decreasing is optimized by a permutation matrix. This class includes not only our Shannon and Fisher objectives but also the nuclear-norm metric of Ref.~\cite{ma2026inverse} ($f(\mu) = \mu^{-1/2}$) and other members of the family $\sum_j \mu_j^{-n/2}$. A detailed proof is given in the SM; we sketch the argument here. Let $\mu_j = \lambda_j^2$ denote the eigenvalues of $\Tv^T\Tv$. The column-sum constraint $\sum_i T_{ij} \le 1$ implies $\|\boldsymbol\mu\|_1 = \mathrm{Tr}[\Tv^T\Tv] \le N$, which in turn places $\boldsymbol\mu$ in its own simplex: $\mu_j \ge 0$, $\|\boldsymbol\mu\|_1 \le N$. Monotonicity of $f$ pushes the optimum to the face $\|\boldsymbol\mu\|_1 = N$; on that face, strict convexity forces all $\mu_j$ to take the same value, which by $\sum_j \mu_j = N$ gives $\boldsymbol\mu = (1,\dots,1)$. Hence $\Tv^T\Tv = I$, and a matrix with nonnegative entries and orthonormal columns must have each column equal to a standard basis vector, implying $\Tv$ is a permutation matrix.

\begin{figure}[tb]
  \centering
  \includegraphics[width=0.9\linewidth]{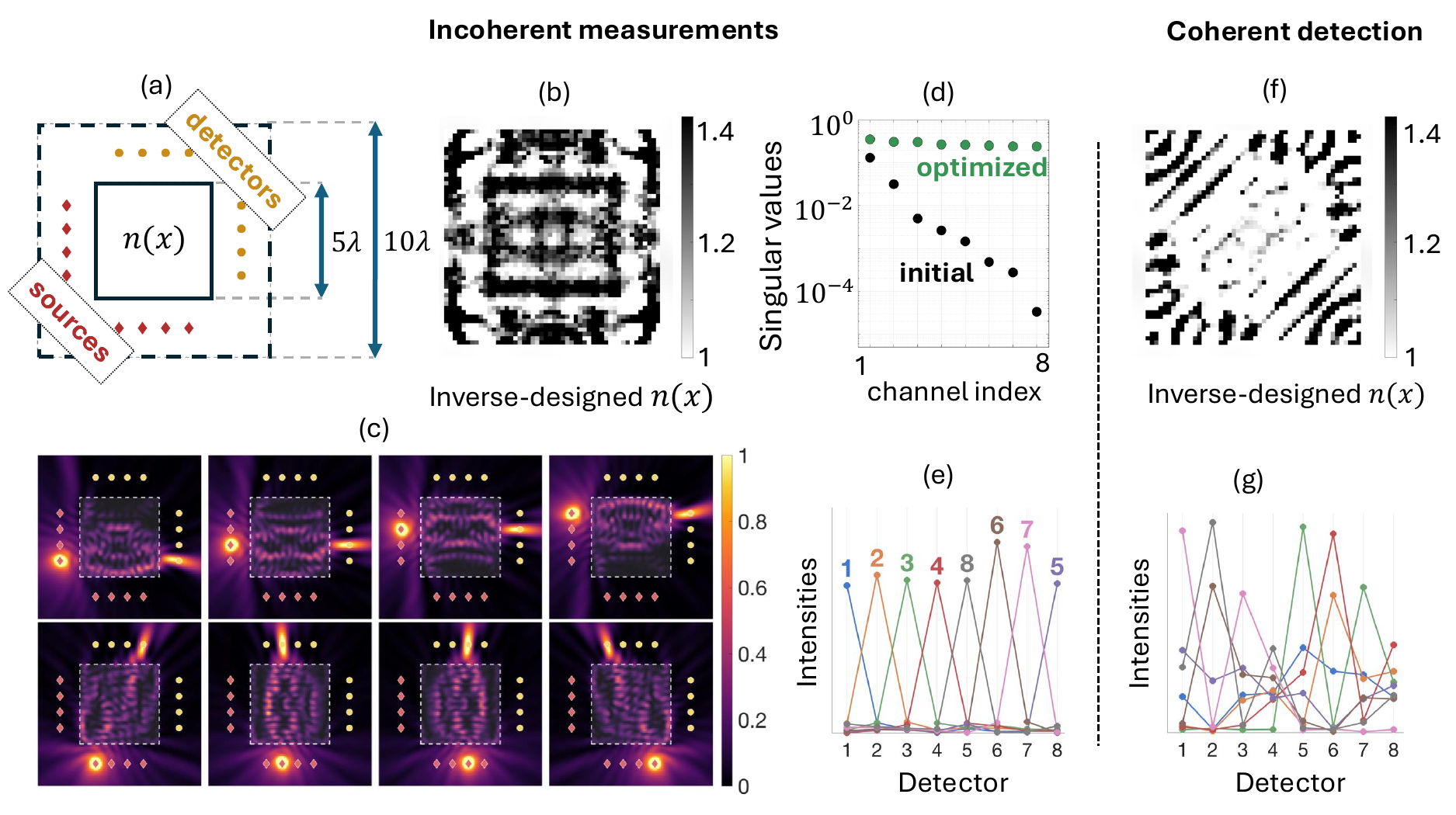}
  \caption{(a) Two-way imager with sources to its left and below (red) and detectors above and to its right (gold). (b)~Optimized refractive index $n(x)$ for incoherent sources with intensity-only detection (Shannon capacity objective). (c)~Field intensities for each of the 8 emitters: the optimized structure routes each source to a distinct detector. (d,e)~Singular values of the incoherent transfer matrix and detector intensities per emitter: after optimization the singular values are nearly equal and the intensity profiles approximate Kronecker deltas, both consistent with a permutation matrix. (f,g)~For comparison, the optimized $n(x)$ and detector intensities for coherent sources with phase-sensitive detection: intensity is spread across all detectors, since any complete basis at the output suffices.}
  \label{fig:two_way_imager}
\end{figure}

\section{Computational Examples: Generalized Focusing and Data-Free Optimization}
\label{sec:CompExamples}
In this section, we computationally validate the two key aspects of the information-theoretic objectives: (1) that they lead to ``generalized focusing'' optical response, and (2) that numerically optimizing with these objectives can produce designs whose performance is comparable (or better than) that of photonic scatterers designed in a full end-to-end pipeline. We consider three examples in which a region of designable refractive index separates source and detector points, and we maximize the information-theoretic objectives. (The gradient derivations for each objective are given in the SM.)  We perform all numerical experiments in 2D (scalar polarization), to alleviate the computational burden of 3D. This should make no difference to the theoretical understanding or comparisons. Each of the three examples represents a setting beyond typical plane-to-plane imaging in free space, and where conventional imaging intuition offers no guidance.

For a first example, we design a ``two-way imager'' that has sources both to its left and below it, with detectors both above and to the right, as shown in Fig.~\ref{fig:two_way_imager}. Our optimization objective is the Shannon information capacity between the eight emitters and eight receivers; notably, the objective treats all emitters and receivers uniformly and does not distinguish between the vertical and horizontal arrays. For wavelength $\lambda$, the designable refractive medium occupies an area of $25\lambda^2$, with the resulting optimal design shown in Fig.~\ref{fig:two_way_imager}(b). The key result is shown in Fig.~\ref{fig:two_way_imager}(c): the optimal index profile finds a way to route the excitations from each source to primarily focus spatially at a single detector. This is reinforced in Fig.~\ref{fig:two_way_imager}(d,e), which show that the optimized singular values are nearly equal and the detector intensities for each source closely approximate the ``Kronecker deltas'' discussed above. To contrast with the case in which one has coherent measurements, \emph{not} intensity measurements, we include a fully coherent case alternative in Fig.~\ref{fig:two_way_imager}(f,g). The design changes, but more importantly the nature of the response changes: now the intensities for any one source are spread across all detectors, as there is no pressure to focus them, because any complete coherent basis at the detectors is suitable.

\begin{figure}[tb]
  \centering
  \includegraphics[width=\linewidth]{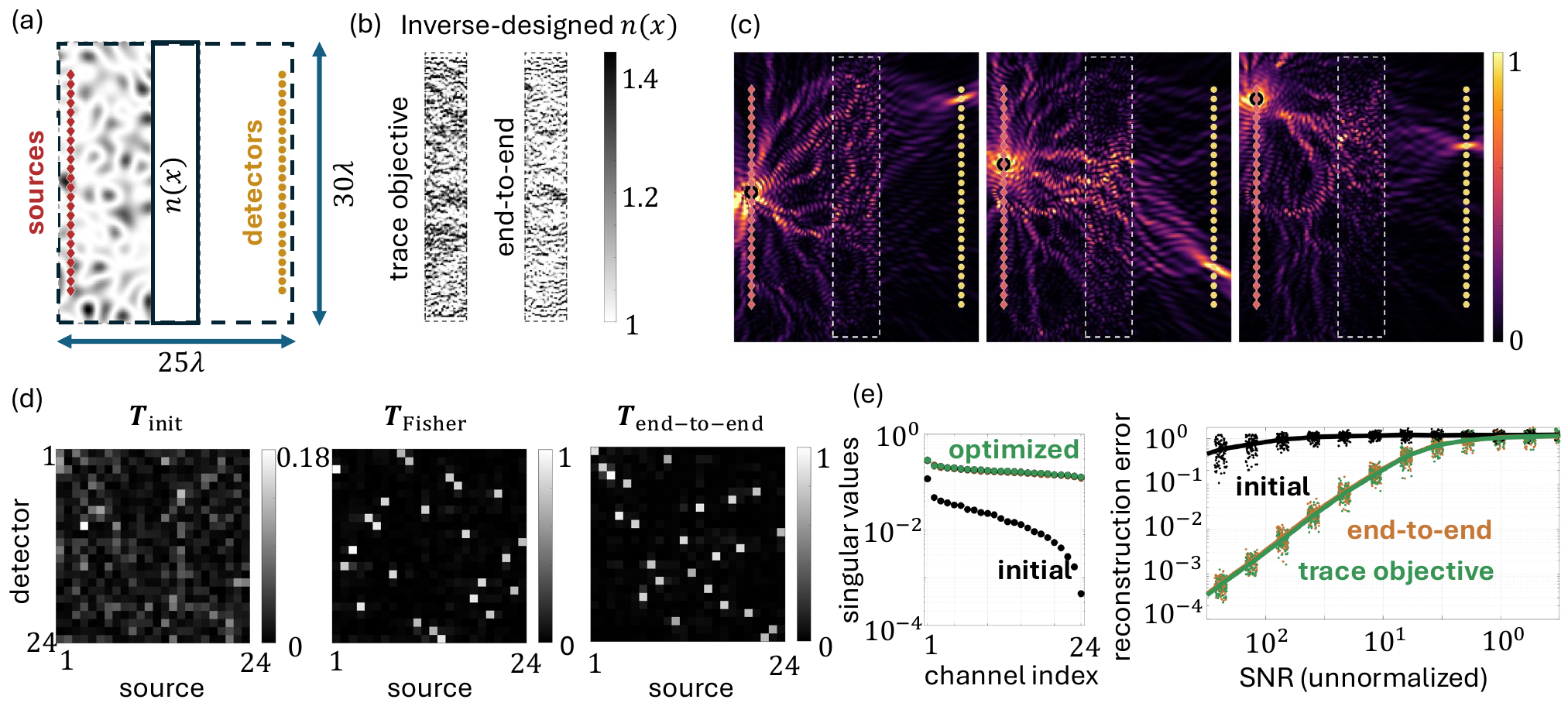}
  \caption{Imaging through a random scattering medium. (a)~Schematic: sources (red) are embedded in a random scatterer; a designable medium $n(x)$ (blue box) sits between the random scatterer and the detector array. (b)~Optimized $n(x)$ for the Fisher objective and for end-to-end optimization. (c)~Field intensities for three representative source excitations, showing that the optimized scatterer routes each speckle pattern to a distinct detector. (d)~Incoherent transfer matrices $\Tv$ before optimization (left), after Fisher-objective optimization (center), and after end-to-end optimization (right); both optimized matrices converge to near-permutation structure. (e)~Singular values before and after optimization (left) and reconstruction error vs.\ SNR (right)---the data-free Fisher objective matches end-to-end performance.}
  \label{fig:random_scatterer}
\end{figure}

As a second example, we consider sources \emph{embedded within a random scattering medium} (Fig.~\ref{fig:random_scatterer}), a scenario motivated by experimental progress in imaging and focusing through disordered media~\cite{vellekoop2007focusing,popoff2010measuring,mosk2012controlling,rotter2017light,cao2022shaping}. (Fisher information and Cram\'er--Rao bounds have separately been used to characterize estimation limits in coherent scattering measurements~\cite{bouchet2021maximum,hupfl2024continuity,starshynov2025model}, though with optimization over input wavefronts rather than over the scattering system itself.) The random scatterer has a broadband reciprocal-space susceptibility (Gaussian noise up to a cutoff $k_{\max} = 0.4k$), producing speckle patterns that impinge upon the designable medium, rather than the spherical or cylindrical waves of free space. We optimize the Fisher objective of Eq.~\ref{eq:FisherObjective}, which for this and all other examples produces essentially equivalent performance to the Shannon objective. The result again confirms generalized focusing: the optimized structure unscrambles the incident speckle, routing each pattern to a distinct detector [Fig.~\ref{fig:random_scatterer}(c)], and the transfer matrix converges to near-permutation form [Fig.~\ref{fig:random_scatterer}(d)].

We also use this example for a direct comparison between our data-free Fisher optimization and a fully implemented end-to-end optimization inspired by Refs.~\cite{lin2021end,lin2022end,sun2021endtoend}. The end-to-end objective targets reconstruction error directly: random ground-truth source patterns $X$ (non-bold denoting a specific instance of $\Xv$; entries i.i.d.\ uniform on $[0,1]$) are generated, noisy measurements simulated, and images reconstructed via Tikhonov-regularized inversion (cf.\ SM):
\begin{equation}
\hat X(\Tv) = (\Tv^T \Tv + \alpha \boldsymbol{I})^{-1}\Tv^T Y,
\end{equation}
where $Y$ denotes the noisy detector intensities. The end-to-end objective is the mean-squared reconstruction error,
\begin{equation}\label{eq:recons_error}
\Big\langle \|X-\hat X(\Tv)\|_2^2 \Big\rangle_{N,X}\ ,
\end{equation}
averaged over $N_n$ noise realizations and $N_g$ randomly drawn ground truths. Gradients of the reconstruction error with respect to the designable medium are computed by back-propagation through the full forward--reconstruct pipeline and iteratively updated, with fresh noise samples drawn at each iteration. In both optimization methods (Fisher and end-to-end), the transfer matrices are computed identically, by running full Maxwell simulations to obtain the intensity pattern generated by each source. The methods differ only in their objective (\eqref{eq:FisherObjective} versus \eqref{eq:recons_error}). To evaluate the two optimized structures, we simulate the reconstruction of a separate, fixed set of $N_g$ ground truths via least-squares reconstruction over this finite set. The reconstruction error is then averaged over the $N_g$ ground truths and $N_n=100$ noise realizations. As shown in Fig.~\ref{fig:random_scatterer}(e), the data-free Fisher objective produces essentially equivalent---if anything, slightly superior---reconstruction performance across a range of SNR values.

\begin{figure}[tb]
  \centering
  \includegraphics[width=0.9\linewidth]{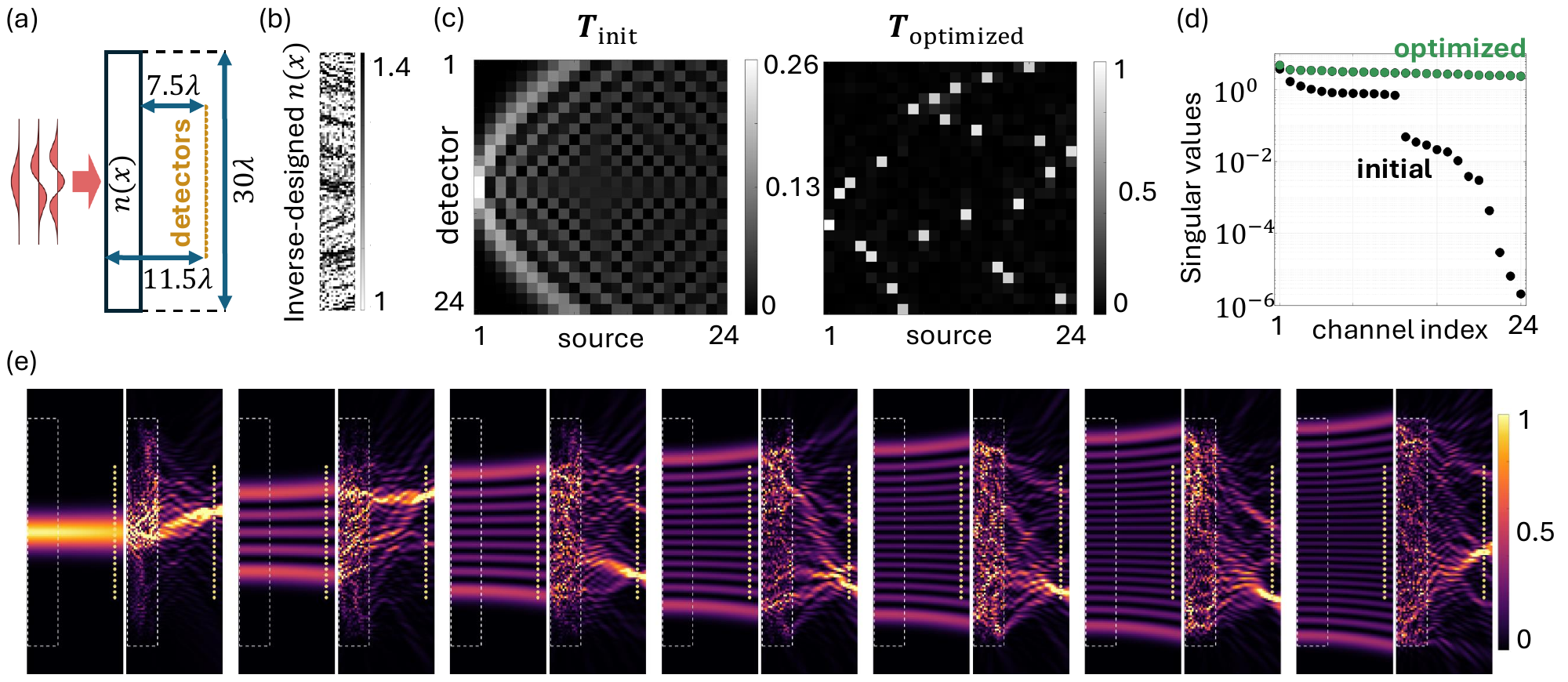}
  \caption{Mode sorting of Hermite--Gauss beams. (a)~Schematic: HG modes impinge from the left onto a designable medium $n(x)$; an array of intensity-only point detectors sits at the output. (b)~Optimized $n(x)$. (c)~Transfer matrices before and after optimization: $T_{\mathrm{init}}$ shows a checkerboard pattern from the alternating symmetry of HG modes; $T_{\mathrm{optimized}}$ is near-permutation. (d)~Singular values before and after optimization. (e)~Field intensities for representative HG modes under the optimized structure, each focusing onto a distinct detector.}
  \label{fig:hermite_gauss}
\end{figure}

As a third example, we consider a signal composed of $N = 24$ collimated Hermite--Gauss (HG) beams that the optics must sort onto an array of $N$ point detectors (Fig.~\ref{fig:hermite_gauss}). Unlike the point-source geometries above, these modes are extended and spatially overlapping, making it far from obvious that a single scattering structure can route each mode to a distinct detector. This scenario connects to mode-division multiplexing and spatial mode sorting, where architectures such as multi-plane light converters~\cite{morizur2010programmable,labroille2014efficient,fontaine2019laguerre} are designed to separate spatial modes into distinct output channels.

We optimize the transfer matrix using the Fisher objective of \eqn{eq:FisherObjective}. Before optimization, the transfer matrix $\Tv_{\mathrm{init}}$ exhibits a checkerboard structure reflecting the alternating spatial symmetry of even- and odd-order HG modes [Fig.~\ref{fig:hermite_gauss}(c)]; the singular values span several orders of magnitude [Fig.~\ref{fig:hermite_gauss}(d)], indicating that most mode information is lost. After optimization, the transfer matrix converges to near-permutation form with nearly equal singular values, and the field intensity plots [Fig.~\ref{fig:hermite_gauss}(e)] confirm that each HG mode is routed to a distinct detector.

A perhaps artificial aspect of this example is the assumption of mutual incoherence between HG modes, which would rarely hold in practice. In the next section, we show that this assumption is in fact unnecessary: generalized-focusing optima persist for coherent and partially coherent inputs alike, so the mode-sorting structure of Fig.~\ref{fig:hermite_gauss} remains optimal regardless of the modes' mutual coherence.

\section{The bottleneck: Intensity measurements, not incoherence}\label{sec:bottleneck}
The linear relation $\Yv = \Tv \Xv$ underlying our analysis assumes perfectly incoherent sources, for which the intensity transfer function is linear. When the input fields are mutually coherent, interference makes the measured intensities a nonlinear function of the source intensities, and the incoherent framework does not directly apply. One might think this nonlinear case cannot be rigorously analyzed, or that the additional degrees of freedom in the coherent field---the relative phases---could be exploited to surpass the generalized-focusing optimum. We show that the opposite is true: for intensity-only detection, non-mixing (permutation-like) transfer is the \emph{only} structure that allows full amplitude recovery, regardless of source coherence.

Consider $N$ coherent input fields $x_j = r_j e^{i\phi_j}$ ($r_j \geq 0$) passing through an $N \times N$ unitary transmission matrix $\Tv$, with intensity-only detection $q_i = |y_i|^2 = |\sum_j T_{ij} x_j|^2$. (It is straightforward to argue that the optimal $\Tv$ among passive, sub-unitary matrices must be unitary; for simplicity, then, we restrict ourselves to the unitary boundary.) The goal is to recover the input amplitudes $r_j$---i.e., the source intensities $r_j^2$---from the measured detector intensities $q_i$, without knowledge of the phases $\phi_j$. This is distinct from phase retrieval~\cite{fienup1982phase,shechtman2015phase}, which seeks to recover the phases (or joint amplitudes and phases) themselves; here we ask the more basic question of whether conventional intensity imaging (source-power recovery) is even possible when the inputs are coherent and only output intensities are measured. Expanding the squared modulus yields
\begin{equation}
  q_i = \sum_{j=1}^N |T_{ij}|^2 r_j^2 + \sum_{j \neq k} T_{ij} T_{ik}^* \, r_j r_k \, e^{i(\phi_j - \phi_k)}.
  \label{eq:coh_intensity}
\end{equation}
The first term contains the desired input powers $r_j^2$, weighted by the element-wise squared modulus $|T_{ij}|^2$. The second term contains coherent interference that depends on the unknown relative phases $\phi_j - \phi_k$. For a phase-independent reconstruction to exist, the interference sum must vanish; for the reconstruction to work for \emph{all} phase configurations, each individual term of the interference sum must vanish:
\begin{equation}
  T_{ij}\, T_{ik}^* = 0, \qquad \forall\, i, \quad \forall\, j \neq k.
  \label{eq:coh_nonmixing}
\end{equation}
Each row of $\Tv$ must therefore contain at most one nonzero element. Since $\Tv$ is unitary, each row has unit norm, so that single nonzero entry must have unit modulus, and the orthogonality between rows requires that these nonzero entries appear in unique columns. Thus, each column also contains exactly one nonzero entry. Hence $\Tv$ must take the form
\begin{equation}
  \Tv = \boldsymbol{D}_{\mathrm{out}}\, \boldsymbol{\Pi}\, \boldsymbol{D}_{\mathrm{in}},
  \label{eq:coh_DPiD}
\end{equation}
where $\boldsymbol{\Pi}$ is a permutation matrix and $\boldsymbol{D}_{\mathrm{out}}, \boldsymbol{D}_{\mathrm{in}}$ are diagonal phase matrices. (Equivalently, the entrywise square $|\Tv|^2$ must itself be a permutation matrix---a vertex of the Birkhoff polytope of doubly stochastic matrices.) For any such $\Tv$, the measured intensities reduce to $q_i = r_{\pi(i)}^2$: a permuted copy of the input powers ($i \rightarrow \pi(i)$), independent of all phases and directly invertible.

A complementary Fisher-information analysis confirms the same conclusion (cf.\ SM): treating the unknown phases as nuisance parameters and marginalizing via the Schur complement of the Fisher matrix, the effective Fisher information for the amplitudes achieves full rank if and only if $\Tv$ satisfies \eqn{eq:coh_DPiD}. 

\begin{figure}[tb]
  \centering
  \includegraphics[width=\linewidth]{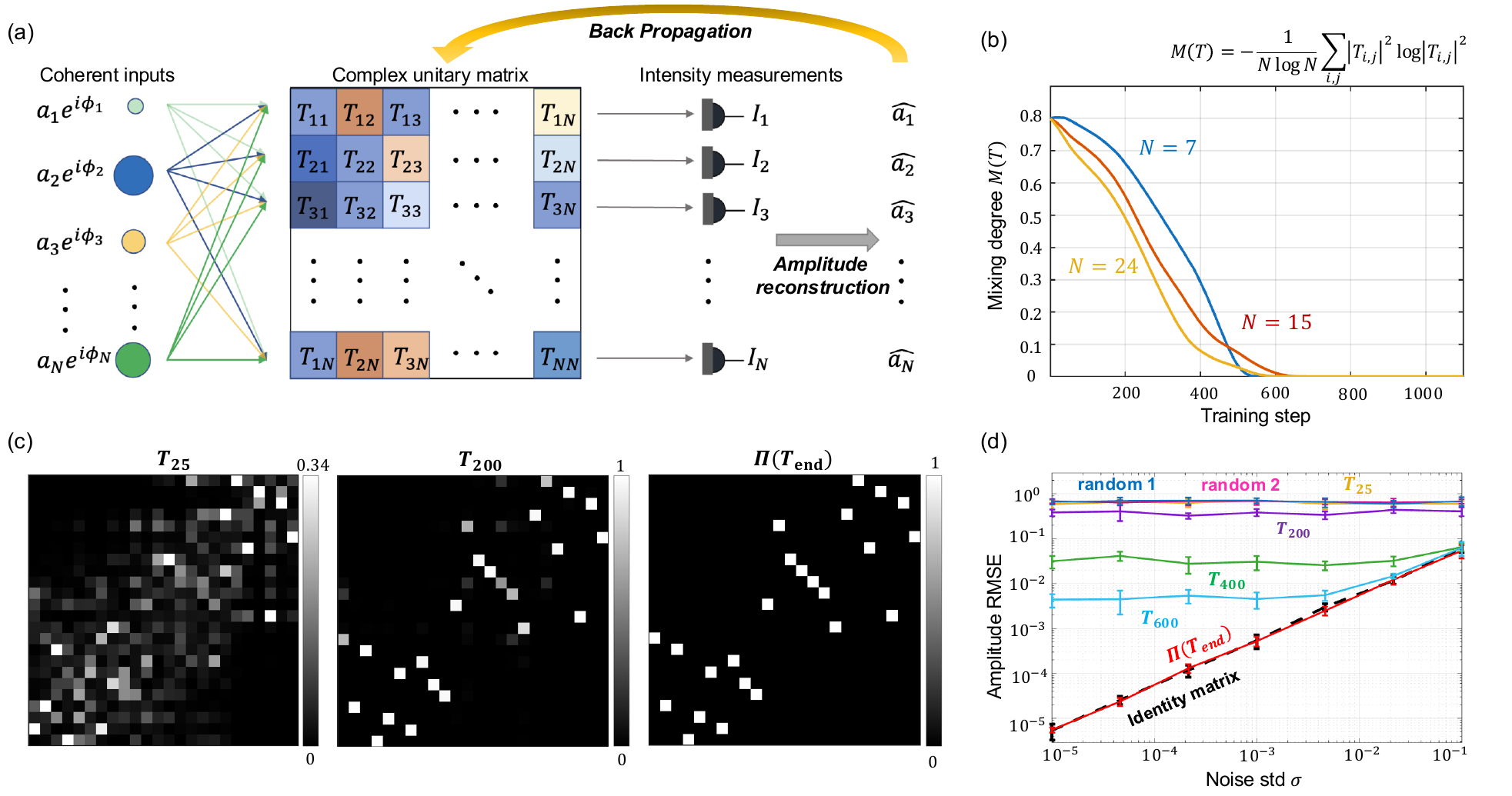}
  \caption{Coherent amplitude retrieval with intensity-only detection. (a)~Schematic: coherent input fields pass through a complex unitary matrix, iteratively optimized to recover input amplitudes with only output intensities measured. (b)~The channel mixing degree decreases toward zero during training. (c, d)~The squared modulus $|T_{ij}|^2$ at training iterations 25, 200, 400, and 600 show the matrix converging towards a permutation matrix. At the final iteration, a final projection onto the nearest permutation matrix (up to complex-valued phases), $\Pi(T_{\rm end})$, yields optimal performance at all tested noise levels.}
  \label{fig:coherent_opt}
\end{figure}

As an independent numerical check, we optimize the unitary matrix directly by gradient-based training (parameterized via Givens rotations to remain on the unitary manifold), minimizing the mean-squared amplitude reconstruction error over random coherent inputs with uniformly distributed amplitudes and phases. Starting from a random initialization, the channel mixing degree---a scalar measure of how far $|\Tv|^2$ is from a permutation matrix---decreases monotonically to zero. The iterations show convergence towards a permutation matrix--the final $|\Tv|^2$ is within $1.5\times 10^{-4}$ of a permutation matrix, as measured in the Frobenius norm, but the convergence is slow and non-trivial at the smallest noise levels. Given the slow convergence, once the mixing degree goes below $10^{-4}$ (between iterations 700 and 900 in our tests), we terminate the iterations and project the amplitudes of $T_{\rm end}$ onto its nearest permutation matrix, which offers optimal performance across all noise levels (Fig.~\ref{fig:coherent_opt}).

The generalized-focusing principle thus extends to coherent sources: under intensity-only detection, the optimal photonic transfer is non-mixing regardless of source coherence. The fundamental constraint is the \emph{intensity bottleneck}---the collapse of complex field amplitudes to nonnegative intensities---not the coherence state of the input.

\section{Discussion}\label{sec:Discussion}
It might have been tempting to think that point-like focusing of waves is no longer important for modern imaging, and that even speckled or random transfer functions can, with back-end training and sufficiently powerful neural networks, produce equally high-resolution images. But our results show that if the scattered waves propagate to intensity-only detectors, then the optimal electromagnetic response is strongly constrained and must correspond to ``generalized focusing.'' Simple heuristics suggest a scaling-law penalty relative to this optimum: if a speckled point-spread function produces $M$ hot spots on average for a single input, then the maximum resolution of the output computational images must be scaled down from order $N$ to order $N/M$. (If $N$ is large enough, $N/M$ can still produce reasonable images, but one has still paid a nontrivial penalty.)

\paragraph*{Connection to conventional imaging.}
In conventional imaging, not only does ``generalized focusing'' occur (in the ideal scenario), but the generalized focusing is \emph{spatially ordered} (neighboring image-plane spots correspond to neighboring source-plane points). We can predict that spatial ordering must arise for fine-grained imagers---those in which planar-arrayed sources and detectors are densely packed at the electromagnetic degrees-of-freedom limit. Consider a source at some position, optimally routed to a particular detector. If we continuously drag that source to a neighboring position, which also routes to a single detector, the smoothness of the Green's function ensures that the optimal detector assignment varies continuously---it cannot jump to a nonlocal detector. In two dimensions, imaging between 1D planes, the only possibilities this allows for are identity or inversion (classic lens) mappings. In three dimensions, imaging between 2D planes, the possibilities expand significantly, as any continuous deformation (rotations, conformal maps) from input to output is allowable. But the most interesting applications of generalized focusing, to us, are beyond the dense, planar-array limits, as demonstrated by the two-way imager, random-scatterer, and Hermite--Gauss imager examples.



\paragraph*{Extensions: continuum sources, unequal channel counts, and noise models.}
In many imaging scenarios, there may not be a natural discretization of a continuous source region into a finite number of points. Electromagnetic degrees of freedom~\cite{toraldo1969degrees,huck1999information,ashok2003information,miller2019waves} prescribe such a conversion: even with an asymptotically large number of source points $S$, a meta-imager cannot support more than $N$ effectively independent source vectors transmitting to $N$ detectors. One could use $S \gg N$ points in a numerical optimization to discover \emph{which} $N$ degrees of freedom a specific scattering scenario is amenable to, or bias the optimization towards a specific set of $N$ modes to be perfectly resolved. (Such trade-offs are familiar in conventional optics; extended-depth-of-field and light-field cameras~\cite{adelson1992single,ng2006digital}, for example, trade lateral resolution for depth resolution, effectively optimizing a different subset of $N$ source ``points'' for efficient information transduction / reconstruction.)

When the number of sources and detectors differs ($S \neq N$), generalized focusing appears to persist in both directions. With \emph{fewer} sources than detectors ($S < N$, so that $\Tv$ is $N \times S$ and $\Tv^T \Tv$ is $S \times S$), the same simplex-and-convexity argument from Sec.~\ref{sec:GenFoc} applies: the eigenvalues of $\Tv^T \Tv$ are bounded by $\operatorname{Tr}[\Tv^T \Tv] \leq S$, convexity pushes them all to 1, and $\Tv^T \Tv$ being diagonal together with non-negativity forces each row to be a distinct canonical basis vector. The requirement that $\Tv^T \Tv = I_S$ imposes further restrictions: The diagonal entries of $\Tv^T \Tv$ are the 2-norms of the columns, which are bounded by the 1-norms, and the 1-norms are bounded by unity. The only way to saturate the inequalities and push all diagonal entries to unity is to make the columns canonical basis vectors, mapping every source to exactly one detector. The surplus of detectors provides only freedom of assignment: any $S$-element subset of detectors can serve as the target, with the remaining $N - S$ detectors unused. With \emph{more} sources than detectors ($S > N$), $\Tv^T\Tv$ has rank at most $N$ and is necessarily singular, so the relevant metrics are $\operatorname{Tr}[(\Tv\Tv^T)^{-1}]$ and $\log\det(\Tv\Tv^T)$, which involve only the $N$ nonzero singular values. The same nonnegativity argument applies: the off-diagonal entries of $\Tv\Tv^T = \sum_j \boldsymbol{t}_j \boldsymbol{t}_j^T$ are sums of nonnegative terms, so the optimal $\Tv\Tv^T \propto \boldsymbol{I}_N$ requires each column to have at most one nonzero entry---still generalized focusing, but now with multiple sources sharing each detector ($S/N$ on average) and only $N$ independent channels resolvable. Recent work on an inverse-designed computational spectrometer~\cite{yu2025spectrometer}, where the number of wavelength channels exceeds the number of sensors, finds an optimized ``banded'' transfer matrix in which each wavelength couples to a small number of adjacent output ports, seemingly at odds with permutation-matrix optimality. We suspect this arises because the photonic structures in those systems (finite-$Q$ resonators with inherent spectral linewidths) simply do not have sufficient degrees of freedom to isolate the response of each wavelength channel to a single sensor.

Our theoretical analysis also assumes additive Gaussian noise, appropriate at high photon counts where read noise dominates or Poisson statistics are well approximated by a Gaussian. At low light levels, shot noise (Poisson statistics) becomes dominant and the Fisher information acquires signal-dependent weights. The generalized-focusing result should extend naturally to this setting, since the core argument rests on the nonnegativity of intensity measurements rather than the specific noise model, but a rigorous treatment is left for future work.

\paragraph*{Prior information.}
Our analysis assumes no prior knowledge of the source distribution: the capacity objective maximizes over all distributions, and the Cram\'er--Rao bound applies to unbiased estimators of arbitrary inputs. When prior information is available---known sparsity, correlations, or restricted support---generalized focusing is likely no longer required. As a simple example, two sources producing non-orthogonal intensity patterns at two detectors can be disentangled if it is known that they never emit simultaneously (sparsity). More broadly, with prior information it is feasible to resolve \emph{more} than $N$ sources with $N$ detectors~\cite{lustig2007sparse,arya2024endtoend}. Recent work on two-point resolution achieves dramatic enhancements through strong prior-knowledge assumptions (e.g., exactly two point emitters on a fixed line)~\cite{tsang2016quantum,nair2016farfield,lupo2016ultimate,paur2016achieving,tham2017beating,tsang2019resolving,wadood2025super}, and extending these ideas to richer priors and more general imager geometries is an important direction. (In such settings, biased estimators---regularized or shrinkage methods~\cite{donoho1994ideal,candes2006robust}---are the natural reconstruction tool; the Cram\'er--Rao bound, which applies to unbiased estimators, is then no longer the relevant performance limit, though the capacity and positivity arguments may still constrain the optics.) One can envision prior-aware analogues of the data-free objectives developed here---replacing the capacity or Fisher metrics with prior-conditioned versions---that could enable closed-form photonic optimization without end-to-end training, even for structured scenes. Alternatively, one can consider nonstatic transfer matrices, jointly optimizing hardware geometry with adaptive measurement policies that condition subsequent acquisitions on intermediate results~\cite{keshvari2026adaptive}.

\paragraph*{Circumventing the bottleneck.}
To go beyond generalized focusing, the intensity bottleneck itself must be overcome: the lack of negative or complex values prevents the construction of nontrivial orthogonal measurement functions. Differential detection---using two nonnegative detector arrays and subtracting their outputs, as employed in some photonic neural network architectures~\cite{wei2024spatially,zheng2024multichannel}---can produce signed values, but only \emph{after} the complex amplitudes have already been collapsed to intensities; information encoded in the phases is still lost at the detectors. A more promising route is interferometric detection at the focal plane~\cite{brady2025interferometric}, which preserves phase information through the measurement process. We encourage continued research in this direction.

\begin{acknowledgments}
We thank A. Douglas Stone and Nazar Pyvovar for helpful conversations. This work was partially supported by the DARPA Coded Visibility STTR program and by the Simons Collaboration on Extreme Wave Phenomena Based on Symmetries (award no. SFI-MPS-EWP-00008530-09).
\end{acknowledgments}

\vspace{5ex}
\textbf{AI Usage:} The great majority of this work was done without any use of AI. As we were finalizing the paper, we used AI in four ways. (1) To help finalize our theory of the coherent-source/incoherent-measurement case. We ``knew'' the right answer, and had a proof for the 2x2 case, and discussions with Claude Opus 4.7 helped quickly generalize our proof to the arbitrary-N case. (2) It helped brainstorm ideas and provide feedback for our schematic figure, Fig. 2. (3) ChatGPT 5.5 augmented our back-propagation codes for Fig. 6 to (slightly) improve our numerical convergence rates. (4) During writing, we used discussions with Claude Opus 4.7 to find logical gaps, discuss points of grammar and style, and develop clarifying language. The (human) authors take full responsibility for all claims made in the paper. The last author has a longstanding fondness for em dashes and declines to scrub them merely to feign the absence of AI collaboration.

\bibliography{refs}

@misc{Lalanne2026-vy,
  title         = {The hidden dimension in nanophotonics design: understanding},
  author        = {Lalanne, P. and Miller, O.},
  month         = apr,
  year          = {2026},
  archivePrefix = {arXiv},
  primaryClass  = {physics.optics},
  eprint        = {2604.07860}
}

@phdthesis{ng2006digital,
  title={Digital light field photography},
  author={Ng, Ren},
  year={2006},
  school={Stanford University}
}

@article{adelson1992single,
  title={Single lens stereo with a plenoptic camera},
  author={Adelson, Edward H and Wang, John YA},
  journal={IEEE Transactions on Pattern Analysis and Machine Intelligence},
  volume={14},
  number={2},
  pages={99--106},
  year={1992},
  publisher={IEEE},
  doi={10.1109/34.121783}
}

@article{wei2024spatially,
  title={Spatially varying nanophotonic neural networks},
  author={Wei, Kaixuan and Li, Xiao and Froech, Johannes and Chakravarthula, Praneeth and Whitehead, James and Tseng, Ethan and Majumdar, Arka and Heide, Felix},
  journal={Science Advances},
  volume={10},
  number={45},
  pages={eadp0391},
  year={2024},
  publisher={American Association for the Advancement of Science},
  doi={10.1126/sciadv.adp0391}
}

@article{zheng2024multichannel,
  title={Multichannel meta-imagers for accelerating machine vision},
  author={Zheng, Hanyu and Liu, Quan and Kravchenko, Ivan I and Zhang, Xiaomeng and Huo, Yuankai and Valentine, Jason G},
  journal={Nature Nanotechnology},
  volume={19},
  pages={471--478},
  year={2024},
  publisher={Nature Publishing Group},
  doi={10.1038/s41565-023-01557-2}
}

@article{brady2025interferometric,
  title={Interferometric focal planes},
  author={Brady, David J and Zhu, Shengtai and Dong, Zhipeng},
  journal={Optics Express},
  volume={33},
  number={10},
  pages={21634--21649},
  year={2025},
  publisher={Optica Publishing Group},
  doi={10.1364/OE.553875}
}

@misc{yu2025spectrometer,
  title         = {Wavelength-scale noise-resistant on-chip spectrometer},
  author        = {Yu, Jianbo and Lo, Hsuan and Chen, Wenduo and Zhu, Changyan
                   and Wu, Yujin and Wang, Fakun and Wang, Chongwu
                   and Yan, Congliao and Dang, Cuong and Wen, Bihan
                   and Cao, Hui and Chong, Yidong and Wang, Qi Jie},
  year          = {2025},
  archivePrefix = {arXiv},
  eprint        = {2509.22286}
}

@article{ma2026inverse,
  title={Inverse Design for Robust Inference in Integrated Computational Spectrometry},
  author={Ma, Wenchao and Pestourie, Rapha{\"e}l and Lin, Zin and Johnson, Steven G},
  journal={Nanophotonics},
  volume={15},
  pages={e70054},
  year={2026},
  publisher={Wiley},
  doi={10.1002/nap2.70054}
}

@article{kuang2025bounds,
  title={Bounds on the coupling strengths of communication channels and their information capacities},
  author={Kuang, Zeyu and Miller, David A B and Miller, Owen D},
  journal={IEEE Transactions on Antennas and Propagation},
  volume={73},
  pages={3959--3974},
  year={2025},
  publisher={IEEE},
  doi={10.1109/TAP.2025.3551975}
}

@article{miller2025tunneling,
  title={Tunneling escape of waves},
  author={Miller, David A B and Kuang, Zeyu and Miller, Owen D},
  journal={Nature Photonics},
  volume={19},
  pages={284--290},
  year={2025},
  publisher={Nature Publishing Group},
  doi={10.1038/s41566-024-01610-9}
}

@article{lustig2007sparse,
  title={Sparse {MRI}: The application of compressed sensing for rapid {MR} imaging},
  author={Lustig, Michael and Donoho, David and Pauly, John M},
  journal={Magnetic Resonance in Medicine},
  volume={58},
  number={6},
  pages={1182--1195},
  year={2007},
  publisher={Wiley},
  doi={10.1002/mrm.21391}
}

@article{donoho1994ideal,
  title={Ideal spatial adaptation by wavelet shrinkage},
  author={Donoho, David L and Johnstone, Iain M},
  journal={Biometrika},
  volume={81},
  number={3},
  pages={425--455},
  year={1994},
  publisher={Oxford University Press},
  doi={10.1093/biomet/81.3.425}
}

@article{candes2006robust,
  title={Robust uncertainty principles: exact signal reconstruction from highly incomplete frequency information},
  author={Cand{\`e}s, Emmanuel J and Romberg, Justin and Tao, Terence},
  journal={IEEE Transactions on Information Theory},
  volume={52},
  number={2},
  pages={489--509},
  year={2006},
  publisher={IEEE},
  doi={10.1109/TIT.2005.862083}
}

@misc{keshvari2026adaptive,
  title         = {Adaptive sensing beyond non-adaptive information limits:
                   End-to-end co-design of geometry, policy, and inference},
  author        = {Keshvari, Arvin and Tuxbury, William and Lin, Zin},
  year          = {2026},
  archivePrefix = {arXiv},
  eprint        = {2604.25193}
}

@article{foschini1998limits,
  title={On limits of wireless communications in a fading environment when using multiple antennas},
  author={Foschini, Gerard J and Gans, Michael J},
  journal={Wireless Personal Communications},
  volume={6},
  number={3},
  pages={311--335},
  year={1998},
  publisher={Springer},
  doi={10.1023/A:1008889222784}
}

@article{fienup1982phase,
  title={Phase retrieval algorithms: a comparison},
  author={Fienup, James R},
  journal={Applied Optics},
  volume={21},
  number={15},
  pages={2758--2769},
  year={1982},
  publisher={Optica Publishing Group},
  doi={10.1364/AO.21.002758}
}

@article{shechtman2015phase,
  title={Phase retrieval with application to optical imaging: a contemporary overview},
  author={Shechtman, Yoav and Eldar, Yonina C and Cohen, Oren and Chapman, Henry N and Miao, Jianwei and Segev, Mordechai},
  journal={IEEE Signal Processing Magazine},
  volume={32},
  number={3},
  pages={87--109},
  year={2015},
  publisher={IEEE},
  doi={10.1109/MSP.2014.2352673}
}

@article{mait2018computational,
    author = {Mait, Joseph N. and Euliss, Gary W. and Athale, Ravindra A.},
    title = {Computational imaging},
    journal = {Adv. Opt. Photon.},
    volume = {10},
    number = {2},
    pages = {409--483},
    year = {2018},
    month = {Jun},
    doi = {10.1364/AOP.10.000409},
}

@article{delbracio2021mobile,
    author = {Delbracio, Mauricio and Kelly, Damien and Brown, Michael S. and Milanfar, Peyman},
    title = {Mobile Computational Photography: A Tour},
    journal = {Annual Review of Vision Science},
    volume = {7},
    pages = {571--604},
    year = {2021},
    doi = {10.1146/annurev-vision-093019-115521},
}

@article{yu2014flat,
    author = {Yu, Nanfang and Capasso, Federico},
    title = {Flat optics with designer metasurfaces},
    journal = {Nature Materials},
    volume = {13},
    number = {2},
    pages = {139--150},
    year = {2014},
    doi = {10.1038/nmat3839},
}

@article{khorasaninejad2016metalenses,
    author = {Khorasaninejad, Mohammadreza and Chen, Wei Ting and Devlin, Robert C. and Oh, Jaewon and Zhu, Alexander Y. and Capasso, Federico},
    title = {Metalenses at visible wavelengths: Diffraction-limited focusing and subwavelength resolution imaging},
    journal = {Science},
    volume = {352},
    number = {6290},
    pages = {1190--1194},
    year = {2016},
    doi = {10.1126/science.aaf6644},
}

@article{chen2020flat,
    author = {Chen, Wei Ting and Zhu, Alexander Y. and Capasso, Federico},
    title = {Flat optics with dispersion-engineered metasurfaces},
    journal = {Nature Reviews Materials},
    volume = {5},
    pages = {604--620},
    year = {2020},
    doi = {10.1038/s41578-020-0203-3},
}

@article{dowski1995extended,
    author = {Dowski, Edward R. and Cathey, W. Thomas},
    title = {Extended depth of field through wave-front coding},
    journal = {Appl. Opt.},
    volume = {34},
    number = {11},
    pages = {1859--1866},
    year = {1995},
    month = {Apr},
    doi = {10.1364/AO.34.001859},
}

@article{fenimore1978coded,
    author = {Fenimore, E. E. and Cannon, T. M.},
    title = {Coded aperture imaging with uniformly redundant arrays},
    journal = {Appl. Opt.},
    volume = {17},
    number = {3},
    pages = {337--347},
    year = {1978},
    month = {Feb},
    doi = {10.1364/AO.17.000337},
}

@article{park2003super,
    author = {Park, Sung Cheol and Park, Min Kyu and Kang, Moon Gi},
    title = {Super-resolution image reconstruction: a technical overview},
    journal = {IEEE Signal Processing Magazine},
    volume = {20},
    number = {3},
    pages = {21--36},
    year = {2003},
    month = {May},
    doi = {10.1109/MSP.2003.1203207},
}

@article{yang2010image,
    author = {Yang, Jianchao and Wright, John and Huang, Thomas S. and Ma, Yi},
    title = {Image Super-Resolution Via Sparse Representation},
    journal = {IEEE Transactions on Image Processing},
    volume = {19},
    number = {11},
    pages = {2861--2873},
    year = {2010},
    month = {Nov},
    doi = {10.1109/TIP.2010.2050625},
}

@article{dong2016image,
    author = {Dong, Chao and Loy, Chen Change and He, Kaiming and Tang, Xiaoou},
    title = {Image Super-Resolution Using Deep Convolutional Networks},
    journal = {IEEE Transactions on Pattern Analysis and Machine Intelligence},
    volume = {38},
    number = {2},
    pages = {295--307},
    year = {2016},
    month = {Feb},
    doi = {10.1109/TPAMI.2015.2439281},
}

@article{levin2007image,
    author = {Levin, Anat and Fergus, Rob and Durand, Fr{\'e}do and Freeman, William T.},
    title = {Image and depth from a conventional camera with a coded aperture},
    journal = {ACM Trans. Graph.},
    volume = {26},
    number = {3},
    articleno = {70},
    year = {2007},
    month = {Jul},
    doi = {10.1145/1276377.1276464},
}

@article{colburn2018metasurface,
    author = {Colburn, Shane and Zhan, Alan and Majumdar, Arka},
    title = {Metasurface optics for full-color computational imaging},
    journal = {Science Advances},
    volume = {4},
    number = {2},
    pages = {eaar2114},
    year = {2018},
    doi = {10.1126/sciadv.aar2114},
}

@article{bayati2022inverse,
    author = {Bayati, Elyas and Pestourie, Rapha{\"e}l and Colburn, Shane and Lin, Zin and Johnson, Steven G. and Majumdar, Arka},
    title = {Inverse designed extended depth of focus meta-optics for broadband imaging in the visible},
    journal = {Nanophotonics},
    volume = {11},
    number = {11},
    pages = {2531--2540},
    year = {2022},
    doi = {10.1515/nanoph-2021-0431},
}

@article{chakravarthula2023thin,
    author = {Chakravarthula, Praneeth and Sun, Jipeng and Li, Xiao and Lei, Chenyang and Chou, Gene and Bijelic, Mario and Froech, Johannes and Majumdar, Arka and Heide, Felix},
    title = {Thin On-Sensor Nanophotonic Array Cameras},
    journal = {ACM Trans. Graph.},
    volume = {42},
    number = {6},
    year = {2023},
    doi = {10.1145/3618398},
}

@article{toraldo1969degrees,
    author = {Toraldo di Francia, G.},
    title = {Degrees of Freedom of an Image},
    journal = {J. Opt. Soc. Am.},
    volume = {59},
    number = {7},
    pages = {799--804},
    year = {1969},
    month = {Jul},
    doi = {10.1364/JOSA.59.000799},
}

@article{tsang2016quantum,
    author = {Tsang, Mankei and Nair, Ranjith and Lu, Xiao-Ming},
    title = {Quantum Theory of Superresolution for Two Incoherent Optical Point Sources},
    journal = {Phys. Rev. X},
    volume = {6},
    issue = {3},
    pages = {031033},
    year = {2016},
    month = {Aug},
    doi = {10.1103/PhysRevX.6.031033},
}

@article{paur2016achieving,
    author = {Pa{\'u}r, Martin and Stoklasa, Bohumil and Hradil, Zdenek and S{\'a}nchez-Soto, Luis L. and Re{\v{h}}{\'a}{\v{c}}ek, Jaroslav},
    title = {Achieving the ultimate optical resolution},
    journal = {Optica},
    volume = {3},
    number = {10},
    pages = {1144--1147},
    year = {2016},
    month = {Oct},
    doi = {10.1364/OPTICA.3.001144},
}

@article{nair2016farfield,
    author = {Nair, Ranjith and Tsang, Mankei},
    title = {Far-Field Superresolution of Thermal Electromagnetic Sources at the Quantum Limit},
    journal = {Phys. Rev. Lett.},
    volume = {117},
    issue = {19},
    pages = {190801},
    year = {2016},
    month = {Nov},
    doi = {10.1103/PhysRevLett.117.190801},
}

@article{lupo2016ultimate,
    author = {Lupo, Cosmo and Pirandola, Stefano},
    title = {Ultimate Precision Bound of Quantum and Subwavelength Imaging},
    journal = {Phys. Rev. Lett.},
    volume = {117},
    issue = {19},
    pages = {190802},
    year = {2016},
    month = {Nov},
    doi = {10.1103/PhysRevLett.117.190802},
}

@article{tham2017beating,
    author = {Tham, Weng-Kian and Ferretti, Hugo and Steinberg, Aephraim M.},
    title = {Beating {R}ayleigh's Curse by Imaging Using Phase Information},
    journal = {Phys. Rev. Lett.},
    volume = {118},
    issue = {7},
    pages = {070801},
    year = {2017},
    month = {Feb},
    doi = {10.1103/PhysRevLett.118.070801},
}

@article{tsang2019resolving,
    author = {Tsang, Mankei},
    title = {Resolving starlight: a quantum perspective},
    journal = {Contemporary Physics},
    volume = {60},
    number = {4},
    pages = {279--298},
    year = {2019},
    doi = {10.1080/00107514.2020.1736375},
}

@misc{wadood2025super,
    author = {Wadood, S. A. and Aarav, Shaurya and Liang, Kevin and Fleischer, Jason W.},
    title = {Super-resolution with {F}ourier measurements},
    year = {2025},
    eprint = {2511.06098},
    archivePrefix = {arXiv},
    primaryClass = {physics.optics},
}

@article{fellgett1955assessment,
    author = {Fellgett, Peter Berners and Linfoot, E. H.},
    title = {On the assessment of optical images},
    journal = {Philosophical Transactions of the Royal Society of London, Series A: Mathematical and Physical Sciences},
    volume = {247},
    number = {931},
    pages = {369-407},
    year = {1955},
    month = {02},
    doi = {10.1098/rsta.1955.0001},
}

@article{linfoot1955information,
    author = {E. H. Linfoot},
    journal = {J. Opt. Soc. Am.},
    number = {10},
    pages = {808--819},
    title = {Information Theory and Optical Images},
    volume = {45},
    month = {Oct},
    year = {1955},
    doi = {10.1364/JOSA.45.000808},
}

@article{ashok2003information,
    author = {Amit Ashok and Mark A. Neifeld},
    journal = {Opt. Express},
    number = {18},
    pages = {2153--2162},
    title = {Information-based analysis of simple incoherent imaging systems},
    volume = {11},
    month = {Sep},
    year = {2003},
    doi = {10.1364/OE.11.002153},
}

@article{huck1999information,
    author = {Huck, Friedrich O. and Fales, Carl L. and Alter-Gartenberg, Rachel and Park, Stephen K. and Rahman, Zia-ur},
    title = {Information-theoretic assessment of sampled imaging systems},
    journal = {Optical Engineering},
    volume = {38},
    number = {5},
    pages = {742--762},
    year = {1999},
    doi = {10.1117/1.602264},
}

@book{cover2006elements,
    author = {Cover, Thomas M. and Thomas, Joy A.},
    title = {Elements of Information Theory},
    edition = {2},
    publisher = {Wiley-Interscience},
    address = {Hoboken, NJ},
    year = {2006},
    isbn = {9780471241959},
}

@book{barrett2004foundations,
    author = {Barrett, Harrison H. and Myers, Kyle J.},
    title = {Foundations of Image Science},
    series = {Wiley Series in Pure and Applied Optics},
    publisher = {Wiley-Interscience},
    address = {Hoboken, NJ},
    year = {2004},
    isbn = {9780471153009},
}

@article{thompson2002precise,
    author = {Thompson, Russell E. and Larson, Daniel R. and Webb, Watt W.},
    title = {Precise nanometer localization analysis for individual fluorescent probes},
    journal = {Biophysical Journal},
    volume = {82},
    number = {5},
    pages = {2775--2783},
    year = {2002},
    month = {May},
    doi = {10.1016/S0006-3495(02)75618-X},
}

@article{ram2006beyond,
    author = {Ram, Sripad and Ward, E. Sally and Ober, Raimund J.},
    title = {Beyond {R}ayleigh's criterion: A resolution measure with application to single-molecule microscopy},
    journal = {Proc. Natl. Acad. Sci. USA},
    volume = {103},
    number = {12},
    pages = {4457--4462},
    year = {2006},
    doi = {10.1073/pnas.0508047103},
}

@article{pavani2009three,
    author = {Pavani, Sri Rama Prasanna and Thompson, Michael A. and Biteen, Julie S. and Lord, Samuel J. and Liu, Na and Twieg, Robert J. and Piestun, Rafael and Moerner, W. E.},
    title = {Three-dimensional, single-molecule fluorescence imaging beyond the diffraction limit by using a double-helix point spread function},
    journal = {Proc. Natl. Acad. Sci. USA},
    volume = {106},
    number = {9},
    pages = {2995--2999},
    year = {2009},
    doi = {10.1073/pnas.0900245106},
}

@article{chao2016fisher,
    author = {Chao, Jerry and Ward, E. Sally and Ober, Raimund J.},
    title = {Fisher information theory for parameter estimation in single molecule microscopy: tutorial},
    journal = {J. Opt. Soc. Am. A},
    volume = {33},
    number = {7},
    pages = {B36--B57},
    year = {2016},
    month = {Jul},
    doi = {10.1364/JOSAA.33.000B36},
}

@article{kabuli2026lensless,
    author = {Kabuli, Leyla A. and Pinkard, Henry and Markley, Eric and Hung, Clara S. and Waller, Laura},
    title = {Designing lensless imaging systems to maximize information capture},
    journal = {Optica},
    volume = {13},
    number = {2},
    pages = {227--238},
    year = {2026},
    month = {Feb},
    doi = {10.1364/OPTICA.570334},
}

@article{stork2008theoretical,
    author = {David G. Stork and M. Dirk Robinson},
    title = {Theoretical foundations for joint digital-optical analysis of electro-optical imaging systems},
    journal = {Appl. Opt.},
    volume = {47},
    number = {10},
    pages = {B64--B75},
    year = {2008},
    month = {Apr},
    doi = {10.1364/AO.47.000B64},
}

@inproceedings{chakrabarti2016learning,
    author = {Chakrabarti, Ayan},
    title = {Learning Sensor Multiplexing Design through Back-propagation},
    booktitle = {Advances in Neural Information Processing Systems (NeurIPS)},
    volume = {29},
    pages = {3081--3089},
    year = {2016},
    eprint = {1605.07078},
    archivePrefix = {arXiv},
}

@article{chang2018hybrid,
    author = {Chang, Julie and Sitzmann, Vincent and Dun, Xiong and Heidrich, Wolfgang and Wetzstein, Gordon},
    title = {Hybrid optical-electronic convolutional neural networks with optimized diffractive optics for image classification},
    journal = {Scientific Reports},
    volume = {8},
    pages = {12324},
    year = {2018},
    doi = {10.1038/s41598-018-30619-y},
}

@article{wetzstein2020inference,
    author = {Wetzstein, Gordon and Ozcan, Aydogan and Gigan, Sylvain and Fan, Shanhui and Englund, Dirk and Solja{\v{c}}i{\'c}, Marin and Denz, Cornelia and Miller, David A. B. and Psaltis, Demetri},
    title = {Inference in artificial intelligence with deep optics and photonics},
    journal = {Nature},
    volume = {588},
    pages = {39--47},
    year = {2020},
    doi = {10.1038/s41586-020-2973-6},
}

@article{barbastathis2019deep,
    author = {Barbastathis, George and Ozcan, Aydogan and Situ, Guohai},
    title = {On the use of deep learning for computational imaging},
    journal = {Optica},
    volume = {6},
    number = {8},
    pages = {921--943},
    year = {2019},
    month = {Aug},
    doi = {10.1364/OPTICA.6.000921},
}

@article{sun2021endtoend,
    author = {Sun, Qilin and Wang, Congli and Fu, Qiang and Dun, Xiong and Heidrich, Wolfgang},
    title = {End-to-end complex lens design with differentiable ray tracing},
    journal = {ACM Trans. Graph.},
    volume = {40},
    number = {4},
    articleno = {71},
    numpages = {13},
    year = {2021},
    doi = {10.1145/3450626.3459674},
}

@article{arya2024endtoend,
    author = {Arya, Gaurav and Li, William F. and Roques-Carmes, Charles and Solja{\v{c}}i{\'c}, Marin and Johnson, Steven G. and Lin, Zin},
    title = {End-to-End Optimization of Metasurfaces for Imaging with Compressed Sensing},
    journal = {ACS Photonics},
    volume = {11},
    number = {5},
    pages = {2077--2087},
    year = {2024},
    doi = {10.1021/acsphotonics.4c00259},
}

@article{sitzmann2018end,
    author = {Sitzmann, Vincent and Diamond, Steven and Peng, Yifan and Dun, Xiong and Boyd, Stephen and Heidrich, Wolfgang and Heide, Felix and Wetzstein, Gordon},
    title = {End-to-end optimization of optics and image processing for achromatic extended depth of field and super-resolution imaging},
    year = {2018},
    publisher = {Association for Computing Machinery},
    volume = {37},
    number = {4},
    doi = {10.1145/3197517.3201333},
    journal = {ACM Trans. Graph.},
    month = jul,
    articleno = {114},
    numpages = {13},
}

@article{tseng2021neural,
    author = {Tseng, Ethan and Colburn, Shane and Whitehead, James and Huang, Luocheng and Baek, Seung-Hwan and Majumdar, Arka and Heide, Felix},
    title = {Neural nano-optics for high-quality thin lens imaging},
    journal = {Nature Communications},
    volume = {12},
    pages = {6493},
    year = {2021},
    doi = {10.1038/s41467-021-26443-0},
}

@article{lin2021end,
    author = {Lin, Zin and Roques-Carmes, Charles and Pestourie, Rapha{\"e}l and Solja{\v{c}}i{\'c}, Marin and Majumdar, Arka and Johnson, Steven G.},
    title = {End-to-end nanophotonic inverse design for imaging and polarimetry},
    journal = {Nanophotonics},
    volume = {10},
    number = {3},
    pages = {1177-1187},
    doi = {10.1515/nanoph-2020-0579},
    year = {2021},
}

@article{lin2022end,
    author = {Zin Lin and Rapha\"{e}l Pestourie and Charles Roques-Carmes and Zhaoyi Li and Federico Capasso and Marin Solja\v{c}i\'{c} and Steven G. Johnson},
    journal = {Opt. Express},
    number = {16},
    pages = {28358--28370},
    title = {End-to-end metasurface inverse design for single-shot multi-channel imaging},
    volume = {30},
    month = {Aug},
    year = {2022},
    doi = {10.1364/OE.449985},
}

@article{telatar1999capacity,
    author = {Telatar, Emre},
    title = {Capacity of Multi-antenna Gaussian Channels},
    journal = {European Transactions on Telecommunications},
    volume = {10},
    number = {6},
    pages = {585-595},
    doi = {10.1002/ett.4460100604},
    year = {1999},
}

@INPROCEEDINGS{Moser2017-wi,
  title     = "Asymptotic capacity results for {MIMO} wireless optical
               communication",
  author    = "Moser, Stefan M and Mylonakis, Michail and Wang, Ligong and
               Wigger, Michele",
  booktitle = "2017 IEEE International Symposium on Information Theory (ISIT)",
  pages     = "536--540",
  year      =  2017,
  doi       = "10.1109/ISIT.2017.8006585"
}

@article{li2020capacity,
    author = {Li, Longguang and Moser, Stefan M. and Wang, Ligong and Wigger, Mich\`{e}le},
    journal = {IEEE Transactions on Information Theory},
    title = {On the Capacity of MIMO Optical Wireless Channels},
    year = {2020},
    volume = {66},
    number = {9},
    pages = {5660-5682},
    doi = {10.1109/TIT.2020.2979716},
}

@article{lapidoth2009capacity,
    author = {Lapidoth, Amos and Moser, Stefan M. and Wigger, Mich\`{e}le A.},
    journal = {IEEE Transactions on Information Theory},
    title = {On the Capacity of Free-Space Optical Intensity Channels},
    year = {2009},
    volume = {55},
    number = {10},
    pages = {4449-4461},
    doi = {10.1109/TIT.2009.2027522},
}

@misc{pinkard2024information,
    author = {Pinkard, Henry and Kabuli, Leyla and Markley, Eric and Chien, Tiffany and Jiao, Jiantao and Waller, Laura},
    title = {Information-driven design of imaging systems},
    year = {2024},
    eprint = {2405.20559},
    archivePrefix = {arXiv},
    primaryClass = {physics.optics},
}

@misc{markley2025ideal,
    author = {Markley, Eric and Pinkard, Henry and Kabuli, Leyla and Singh, Nalini and Waller, Laura},
    title = {Computationally Efficient Information-Driven Optical Design with Interchanging Optimization},
    year = {2025},
    eprint = {2507.07789},
    archivePrefix = {arXiv},
    primaryClass = {eess.IV},
}

@book{janesick2007photon,
    author = {Janesick, James R.},
    title = {Photon Transfer},
    series = {SPIE Press Monograph},
    volume = {PM170},
    publisher = {SPIE Press},
    address = {Bellingham, WA},
    year = {2007},
    isbn = {9780819467225},
    doi = {10.1117/3.725073},
}

@article{healey1994radiometric,
    author = {Healey, Glenn E. and Kondepudy, Raghava},
    title = {Radiometric {CCD} camera calibration and noise estimation},
    journal = {IEEE Transactions on Pattern Analysis and Machine Intelligence},
    volume = {16},
    number = {3},
    pages = {267--276},
    year = {1994},
    doi = {10.1109/34.276126},
}

@article{shechtman2014optimal,
    title = {Optimal Point Spread Function Design for 3D Imaging},
    author = {Shechtman, Yoav and Sahl, Steffen J. and Backer, Adam S. and Moerner, W. E.},
    journal = {Phys. Rev. Lett.},
    volume = {113},
    issue = {13},
    pages = {133902},
    numpages = {5},
    year = {2014},
    month = {Sep},
    publisher = {American Physical Society},
    doi = {10.1103/PhysRevLett.113.133902},
}

@article{vellekoop2007focusing,
    author = {I. M. Vellekoop and A. P. Mosk},
    journal = {Opt. Lett.},
    number = {16},
    pages = {2309--2311},
    title = {Focusing coherent light through opaque strongly scattering media},
    volume = {32},
    month = {Aug},
    year = {2007},
    doi = {10.1364/OL.32.002309},
}

@article{cao2022shaping,
    author = {Cao, Hui and Mosk, Allard P. and Rotter, Stefan},
    title = {Shaping the propagation of light in complex media},
    journal = {Nature Physics},
    volume = {18},
    pages = {994--1007},
    year = {2022},
    doi = {10.1038/s41567-022-01677-x},
}

@article{miller2013self,
    author = {Miller, David A. B.},
    title = {Self-configuring universal linear optical component},
    journal = {Photonics Research},
    volume = {1},
    number = {1},
    pages = {1--15},
    year = {2013},
    doi = {10.1364/PRJ.1.000001},
}

@article{miller2019waves,
    author = {Miller, David A. B.},
    title = {Waves, modes, communications, and optics: a tutorial},
    journal = {Adv. Opt. Photon.},
    volume = {11},
    number = {3},
    pages = {679--825},
    year = {2019},
    month = {Sep},
    doi = {10.1364/AOP.11.000679},
}

@article{morizur2010programmable,
    author = {Morizur, Jean-Fran{\c{c}}ois and Nicholls, Lachlan and Jian, Pu and Armstrong, Seiji and Treps, Nicolas and Hage, Boris and Hsu, Magnus and Bowen, Warwick and Janousek, Jiri and Bachor, Hans-A.},
    title = {Programmable unitary spatial mode manipulation},
    journal = {J. Opt. Soc. Am. A},
    volume = {27},
    number = {11},
    pages = {2524--2531},
    year = {2010},
    month = {Nov},
    doi = {10.1364/JOSAA.27.002524},
}

@article{labroille2014efficient,
    author = {Labroille, Guillaume and Denolle, Bertrand and Jian, Pu and Genevaux, Philippe and Treps, Nicolas and Morizur, Jean-Fran{\c{c}}ois},
    title = {Efficient and mode selective spatial mode multiplexer based on multi-plane light conversion},
    journal = {Opt. Express},
    volume = {22},
    number = {13},
    pages = {15599--15607},
    year = {2014},
    month = {Jun},
    doi = {10.1364/OE.22.015599},
}

@article{fontaine2019laguerre,
    author = {Fontaine, Nicolas K. and Ryf, Roland and Chen, Haoshuo and Neilson, David T. and Kim, Kwangwoong and Carpenter, Joel},
    title = {Laguerre-{G}aussian mode sorter},
    journal = {Nature Communications},
    volume = {10},
    pages = {1865},
    year = {2019},
    doi = {10.1038/s41467-019-09840-4},
}

@article{popoff2010measuring,
    title = {Measuring the Transmission Matrix in Optics: An Approach to the Study and Control of Light Propagation in Disordered Media},
    author = {Popoff, S. M. and Lerosey, G. and Carminati, R. and Fink, M. and Boccara, A. C. and Gigan, S.},
    journal = {Phys. Rev. Lett.},
    volume = {104},
    issue = {10},
    pages = {100601},
    numpages = {4},
    year = {2010},
    month = {Mar},
    publisher = {American Physical Society},
    doi = {10.1103/PhysRevLett.104.100601},
}

@article{mosk2012controlling,
    author = {Mosk, Allard P. and Lagendijk, Ad and Lerosey, Geoffroy and Fink, Mathias},
    title = {Controlling waves in space and time for imaging and focusing in complex media},
    journal = {Nature Photonics},
    volume = {6},
    pages = {283--292},
    year = {2012},
    doi = {10.1038/nphoton.2012.88},
}

@article{rotter2017light,
    author = {Rotter, Stefan and Gigan, Sylvain},
    title = {Light fields in complex media: Mesoscopic scattering meets wave control},
    journal = {Rev. Mod. Phys.},
    volume = {89},
    issue = {1},
    pages = {015005},
    year = {2017},
    doi = {10.1103/RevModPhys.89.015005},
}

@article{bouchet2021maximum,
    author = {Bouchet, Dorian and Rotter, Stefan and Mosk, Allard P.},
    title = {Maximum information states for coherent scattering measurements},
    journal = {Nature Physics},
    volume = {17},
    pages = {564--568},
    year = {2021},
    doi = {10.1038/s41567-020-01137-4},
}

@article{hupfl2024continuity,
    author = {H{\"u}pfl, Jakob and Russo, Felix and Rachbauer, Lukas M. and Bouchet, Dorian and Lu, Junjie and Kuhl, Ulrich and Rotter, Stefan},
    title = {Continuity equation for the flow of {F}isher information in wave scattering},
    journal = {Nature Physics},
    volume = {20},
    number = {8},
    pages = {1294--1299},
    year = {2024},
    doi = {10.1038/s41567-024-02519-8},
}

@article{starshynov2025model,
    author = {Starshynov, Ilya and Weimar, Maximilian and Rachbauer, Lukas M. and Hackl, G{\"u}nther and Faccio, Daniele and Rotter, Stefan and Bouchet, Dorian},
    title = {Model-free estimation of the {C}ram{\'e}r--{R}ao bound for deep learning microscopy in complex media},
    journal = {Nature Photonics},
    volume = {19},
    pages = {593--600},
    year = {2025},
    doi = {10.1038/s41566-025-01657-6},
}

\end{document}